\documentclass[twocolumn,pre,showpacs]{revtex4}

\usepackage{graphicx}
\usepackage{amsmath}
\usepackage{amssymb}
\usepackage{amsfonts}
\usepackage{bm}

\begin{document}

\title{Scaling properties of the asymmetric exclusion process with long-range hopping}

\author{J. Szavits-Nossan}
\email{juraj@ifs.hr}
\author{K. Uzelac}
\email{katarina@ifs.hr}

\affiliation{Institute of Physics, Bijeni\v{c}ka cesta 46, P.O.Box 304, HR-10001 Zagreb, Croatia}

\begin{abstract}
The exclusion process in which particles may jump any distance $l\geq 1$ with the probability that decays as $l^{-(1+\sigma)}$ is studied from coarse-grained equation for density profile in the limit when the lattice spacing goes to zero. For $1<\sigma<2$, the usual diffusion term of this equation is replaced by the fractional one, which affects dynamical-scaling properties of the late-time approach to the stationary state. When applied to an open system with totally asymmetric hopping, this approach gives two results: first, it accounts for the $\sigma$-dependent exponent that characterizes the algebraic decay of density profile in the maximum-current phase for $1<\sigma<2$, and second, it shows that in this region of $\sigma$ the exponent is of the mean-field type.

\end{abstract}


\date{\today}

\maketitle

\section{Introduction}

The asymmetric exclusion process (ASEP) is one of the simplest models for transport in which the net flow of particles is maintained by contact with two reservoirs at different densities $\rho_L$ and $\rho_R$. It emerges in the wide range of non-equilibrium phenomena, like biological transport \cite{MacDonaldGibbsPipkin68, MacDonaldGibbsPipkin69}, surface growth \cite{MeakinRamanlalSanderBall86,KimKosterlitz89} and traffic flow \cite {NagelSchreckenberg92}. In the presence of the net current flowing through the system, the boundary conditions generally play an important role in determining the bulk properties \cite{Krug91}. The phenomenological domain-wall approach \cite{Kolomeisky98} and the exact solution \cite{SchutzDomany93,DerridaEvans93} revealed that the model exhibits phase transitions both of the first and the second order. Description of these new phenomena contributed to considerable efforts taken in the last several decades in order to determine the nature of phase transitions in system held far from equilibrium. Important aspect of these efforts concerns a better understanding of those parameters (e.g. symmetry of the Hamiltonian, dimensionality and range of the interactions) that usually determine an underlying universality among various models of equilibrium phase transitions, but remain important for out-of-equilibrium phase transitions as well (for a recent review on universality classes in non-equilibrium lattice models, see \cite{Odor04}).

In that spirit, one of striking features of ASEP is the robustness of its phase diagram to various modifications. Among a number of different extensions of this model that have been proposed, the universal character \cite{Krug91,JanssenOerding96,HagerKrugPopkovSchutz01} of the continuous phase transition has been verified in a number of cases, including ASEP with parallel update \cite{EvansRajewskiSpeer99,deGierNienhuis99}, partially asymmetric exclusion process \cite{Sasamoto00} and particle-wise disorder \cite{Bengrine99}. With the purpose to examine the impact of the long-range hopping on the character of the phase transitions, a generalized model was introduced \cite{SzavitsUzelac06} in which particles may jump any distance $l\geq 1$ with the probability that decays as $l^{-(1+\sigma)}$. Due to the long range of hopping, the exchange of particles with reservoirs is possible at each site of the lattice, which can be compared to the system with a bulk reservoir \cite{ParmeggianiFranoschFrey03,ParmeggianiFranoschFrey04}. For $\sigma>1$, this generalized model has the same phase diagram as the short-range case consisting of the low-density, high-density and the maximum-current phase, but different effects are found at the transition lines. Besides the localization of the domain-wall at first-order phase transition, the continuous phase transition to the maximum-current phase is accompanied by the exponent that differs from the short-range value $1/2$ in the region $1<\sigma<2$, where its dependence on $\sigma$ was given by the conjecture based on numerical simulations \cite{SzavitsUzelac06}.
 
In the present work we show that the conjectured expression for the $\sigma$-dependent exponent can be obtained in the mean-field approximation. We consider the hydrodynamic approach that gives coarse-grained equation for density profile and apply it to an open system. Within the same formalism we examine dynamical scaling properties of this model in more general asymmetric and symmetric cases with periodic boundary conditions and confirm the predictions by numerical simulations.

The work is organized as follows. In Section II we consider the process on the infinite lattice and find the continuous limit of lattice equations in a general case including both the symmetric and the asymmetric exclusion process. In Section III we analyze the effects of introducing the boundary conditions that correspond to those of the maximum-current phase and 
deduce the analytical expression for the $\sigma$-dependent exponent. Direct numerical solution of lattice equations in the mean-field approximation is also given. The Section IV is dedicated to the analysis of the relaxation to the stationary state. A brief summary of results is given in Section V.

\section{Hydrodynamic approach}

Let us first consider the exclusion process on the infinite one-dimensional lattice, where each site $n$ is either occupied by a particle ($\tau_n=1$) or empty ($\tau_n=0$). Dynamics of this process is described as follows. At any given time $t$, a randomly chosen particle at site $n$ attempts to jump either to the site $n-l$ with a probability $q$, or to the site $n+l$ with a probability $p=1-q$. Distance $l>0$ is chosen according to the probability distribution $p_{l}=l^{-(1+\sigma)}/\zeta(\sigma+1)$, where $\zeta(z)$ is the Riemann zeta function. Generally, one has $p\neq q$ (the asymmetric exclusion process), or $p=q$ (the symmetric exclusion process). In the limit $\sigma\rightarrow\infty$ the hopping reduces only to nearest neighbors, and the process is identical to the exclusion process with short-range hopping. 

This dynamics corresponds to a continuous-time Markov process in which the probability $P(C,t)$ of the system being in a state $C=\{\tau_n\}_{n\in\mathbb{Z}}$ at time $t$ evolves according to the following master equation 

\begin{equation}\label{master}
\mbox{\fontsize{9}{10.8}\selectfont $\displaystyle \frac{\partial P(C,t)}{\partial t}=\sum_{C'}{W(C'\rightarrow C)P(C',t)}-\sum_{C'}{W(C\rightarrow C')P(C,t)}$,}
\end{equation}

\noindent where $W(C\rightarrow C')$ is the transition probability per unit time for a system to go from the state $C$ to the state $C'$. In the model considered here, the non-diagonal elements $W(C\rightarrow C')$ are equal to $q\cdot p_{n-m}$ if $C'=C^{(n,m)}$ and $m<n$, or to $p\cdot p_{m-n}$ if $C'=C^{(m,n)}$ and $m>n$, where $C^{(n,m)}$ is the configuration obtained from $C$ by moving a particle from a site $n$ to an empty site $m$ (if possible).

\noindent Starting from the master equation (\ref{master}), lattice equation for the time-dependent \emph{average} density profile $\langle\tau_n\rangle(t)\equiv\sum_{C}\tau_n P(C,t)$ reads

\begin{equation}\label{tau_n}
\frac{d}{dt}\langle\tau_n\rangle(t)=\langle K_{n}^{(1)}\rangle,
\end{equation}

\noindent where 

\begin{eqnarray}\label{K_n}
K_n^{(1)}&=&\sum_{r>0}p_r(\Delta_r^{+}\tau_n-\Delta_r^{-}\tau_n)-{}\nonumber\\
&&-{}(p-q)\sum_{r>0}p_r\left[(1-\tau_n)\Delta_r^{+}\tau_n+\tau_n\Delta_r^{-}\tau_n\right]
\end{eqnarray}

\noindent and the following notation has been used, $\Delta_r^{+}\tau_n\equiv\tau_{n+r}-\tau_n$ and $\Delta_r^{-}\tau_n\equiv\tau_{n}-\tau_{n-r}$. 

Basically, hydrodynamic equation that we are interested in can be obtained by a suitable coarse-grained procedure when the lattice spacing $a$ goes to zero. This is usually referred as the hydrodynamic limit and describes the time evolution of a system at large time and space scales where the stochastic details of the particular process have been smoothened out. So far, rigorous results have been obtained in the short-range \cite{BenassiFouque87, Rezakhanlou91} and recently in the long-range case for $p=q$ \cite{Jara07}. In the short-range case, the equation for the coarse-grained density $\rho(x,t)$ is given by the inviscid Burgers' equation in the asymmetric case $p\neq q$,

\begin{equation}\label{burg}
\frac{\partial\rho}{\partial t}=-(p-q)\frac{\partial}{\partial x}[\rho(1-\rho)]
\end{equation}

\noindent and by the normal diffusion equation in the symmetric case $p=q$,

\begin{equation}\label{dif}
\frac{\partial\rho}{\partial t}=\frac{1}{2}\frac{{\partial}^2\rho}{\partial x^2}.
\end{equation}

\noindent In the long-range case \cite{Jara07}, diffusion equation (\ref{dif}) is replaced by space-fractional diffusion equation in the region $1<\sigma<2$,

\begin{equation}\label{fdif}
\frac{\partial\phi}{\partial t}=\nu_{\sigma}\Delta_\sigma\phi(x,t),\quad 1<\sigma<2, 
\end{equation}

\noindent where $\Delta_{\sigma}$ is fractional Laplacian (\ref{Delta}) and $\nu_{\sigma}=-2p\Gamma(-\sigma)cos(\pi\sigma/2)/\zeta(\sigma+1)>0$. 

\indent To study the asymmetric case we adopt the non-rigorous approach that consists of taking the mean-field approximation $\langle\tau_n\tau_m\rangle\rightarrow\langle\tau_n\rangle\langle\tau_m\rangle$, $n\neq m$ and applying it directly to lattice equations (\ref{tau_n}) and (\ref{K_n}). Notice that the same approach, when applied to the short-range case, gives the correct hydrodynamic limit although the starting mean-field assumption is only approximate (in the symmetric case, the assumption is not necessary since equation is linear). The reason for this lies in the fact that for certain initial profiles stationary state becomes a product measure (i.e. a factorized state without correlations), so in this case the mean-field approximation becomes exact. It is worth mentioning that the factorized state is preserved even upon the introduction of long-range hopping \cite{Liggett77}, provided that a hopping probability has a finite mean (in case $p_l\sim l^{-(1+\sigma)}$, this is true for $\sigma>1$).

\indent Applying the mean-field approximation to lattice equations (\ref{tau_n}) and (\ref{K_n}), we obtain the following equations 

\begin{eqnarray}\label{phi_inf}
\frac{d\phi_n}{dt}&=&\sum_{r>0}\frac{p_r}{2}(\Delta_r^{+}\phi_n-\Delta_r^{-}\phi_n)+{}\nonumber\\
&&+{}(\Delta\rho+\phi_n)(p-q)\sum_{r>0}p_r(\Delta_r^{+}\phi_n+\Delta_r^{-}\phi_n)
\end{eqnarray}

\noindent where $\phi_n(t)=\langle\tau_n\rangle(t)-\bar{\rho}$ denotes deviation from the uniform profile of density $\bar{\rho}$ and $\Delta\rho=\bar{\rho}-1/2$ is introduced in order to distinguish two cases, $\bar{\rho}=1/2$ ($\Delta\rho=0$) and $\bar{\rho}\neq 1/2$ ($\Delta\rho\neq 0$). To obtain equation for the macroscopic profile $\phi(x,t)$ in the continuous limit, we adopt the procedure given in \cite{Tarasov06}. Basically, one assumes that $\phi_n(t)$ are coefficients in the Fourier series of some function $\hat{\phi}(k,t)$ defined on $[-K/2,K/2]$

\begin{equation}\label{phi_hat}
\hat{\phi}(k,t)=\sum_{n=-\infty}^{\infty}\phi_n(t)e^{-ikx_n},
\end{equation}

\noindent where $x_n=na$ and $a=2\pi/K$ is lattice constant. Then, Eq. (\ref{phi_inf}) can be written in the Fourier space

\begin{eqnarray}\label{phi_inf2}
\frac{d}{dt}\hat{\phi}(k,t)&=&\hat{\phi}(k,t)[D(ka)-D(0)]+\Delta\rho\hat{\phi}(k,t)B(ka)+{}{}\nonumber\\
&& {}+\frac{1}{K^2}\int_{-K/2}^{K/2}dk_1\int_{-K/2}^{K/2}dk_2\hat{\phi}(k_1,t)\hat{\phi}(k_2,t)\cdot\nonumber\\
&& {}\cdot\sum_{n=-\infty}^{\infty}e^{i(k_1+k_2-k)na}B(ka).
\end{eqnarray}

\noindent where $D(ka)$ and $B(ka)$ are given by

\begin{subequations}
\begin{equation}\label{D}
D(ka)=\frac{1}{2}\left[Li_{\sigma+1}(e^{ika})+Li_{\sigma+1}(e^{-ika})\right],
\end{equation}

\begin{equation}\label{B}
B(ka)=(p-q)[Li_{\sigma+1}(e^{ika})-Li_{\sigma+1}(e^{-ika})]
\end{equation}
\end{subequations}

\noindent and $Li_{\nu}(z)$ is the polylogarithm function. Using the series representation of the polylogarithm function valid for non-integer $\nu\neq 1,2,3\dots$ \cite{Robinson51}, one obtains

\begin{widetext}
\begin{subequations}
\begin{equation}\label{D2}
D(ka)-D(0)=\frac{1}{2\zeta(\sigma+1)}\left[2\Gamma(-\sigma)cos\frac{\pi\sigma}{2}\vert k\vert^{\sigma}a^{\sigma}
+2\sum_{n=1}^{\infty}\frac{\zeta(\sigma+1-2n)}{(2n)!}(ik)^{2n}a^{2n}\right],
\end{equation}

\begin{equation}\label{B2}
B(ka)=\frac{p-q}{\zeta(\sigma+1)}\left[-2i\Gamma(-\sigma)sin\frac{\pi\sigma}{2}sgn(k)\vert k\vert^{\sigma}a^{\sigma}
+2\sum_{n=1}^{\infty}\frac{\zeta(\sigma+2-2n)}{(2n-1)!}(ik)^{2n-1}a^{2n-1}\right].
\end{equation}
\end{subequations}
\end{widetext}

\noindent Finally, one defines the macroscopic density profile $\phi(x,t)$ as the inverse Fourier transform of $\tilde{\phi}(k,t)$, obtained from $\hat{\phi}(k,t)$ in the limit $a\rightarrow 0$ after the appropriate scaling $t\rightarrow t/a^z$ has been taken, where $z$ is the lowest exponent in $a$ in (\ref{D2}) and (\ref{B2}). The latter should not be confused with dynamical exponent $z$ considered in Section IV.

\indent \textit{The symmetric case ($p=q$).} From (\ref{D2}) it follows that $z=min\{\sigma,2\}$. For $\sigma>2$ under diffusive scaling ($z=2$) one obtains the normal diffusion equation

\begin{equation}\label{norm_diff}
\frac{\partial\phi}{\partial t}=\nu_{2}\frac{\partial^2\phi}{\partial x^2},\quad \sigma>2,
\end{equation}

\noindent with the diffusion coefficient $\nu_{2}=\zeta(\sigma-1)/2\zeta(\sigma+1)>0$. On the other hand, in the region $1<\sigma<2$ one obtains the \textit{space-fractional} diffusion equation

\begin{equation}\label{frac_diff}
\frac{\partial\phi}{\partial t}=\nu_{\sigma}\Delta_\sigma\phi,\quad 1<\sigma<2, 
\end{equation}

\noindent where $\nu_{\sigma}=-\Gamma(-\sigma)cos(\pi\sigma/2)/\zeta(\sigma+1)>0$, as in Ref. \cite{Jara07}.

\indent \textit{The asymmetric case ($p\neq q$).} From (\ref{D2}) and (\ref{B2}) it follows that $z=min\{\sigma,1\}$. For $\sigma>1$ under Eulerian scaling ($z=1$) this gives the inviscid Burgers' equation with the additional drift term $-v\partial/\partial x\phi(x,t)$

\begin{equation}\label{burgers_drift_sigma}
\frac{\partial\phi}{\partial t}=-v\frac{\partial\phi}{\partial x}-\kappa\phi\frac{\partial\phi}{\partial x},
\quad \sigma>1,
\end{equation}

\noindent where the collective velocity $v$ and $\kappa$ are given by

\begin{equation}\label{v}
v=(p-q)(1-2\bar{\rho})\lambda(\sigma)=\left.\frac{dj(\rho)}{d\rho}\right\vert_{\rho=\bar{\rho}},
\end{equation}

\begin{equation}\label{kappa}
\kappa=-2(p-q)\lambda(\sigma)=\left.\frac{d^2j(\rho)}{d{\rho}^2}\right\vert_{\rho=\bar{\rho}}.
\end{equation}

\noindent In the above expressions, $\lambda(\sigma)$ is the average hopping length and $j(\rho)$ is the macroscopic
current

\begin{equation}
j(\rho)=(p-q)\lambda(\sigma)\rho(1-\rho),\quad \lambda(\sigma)=\frac{\zeta(\sigma)}{\zeta(\sigma+1)}.
\end{equation}

\noindent The original inviscid Burgers' equation (\ref{burg}) is then recovered either by Galilean transformation $x\rightarrow x-vt$ or by taking $\bar{\rho}=1/2$.

We can go further and take a look at the next higher-order terms in (\ref{D2}) and (\ref{B2}). For $1<\sigma<2$ there are two terms, both proportional to $a^{\sigma}$. The first one follows from (\ref{D2}) and corresponds to the fractional Laplacian $\Delta_{\sigma}\phi$ (also known as the Riesz fractional derivative), while the other one is nonlinear and reads $\phi H_{\sigma}\phi$, where $H_{\sigma}$ is defined in (\ref{H}). For $\sigma>2$, the next term in (\ref{D2}) corresponds to the usual diffusion term term $\Delta\phi(x)$ and is proportional to $a$. If we neglect the nonlinear term $\phi H_{\sigma}\phi$, we obtain following equations

\begin{equation}\label{burg_reg_frac}
\frac{\partial\phi^a}{\partial t}=a^{\sigma-1}\nu_{\sigma}\Delta_{\sigma}\phi^a-\kappa\phi^a\frac{\partial\phi^a}{\partial x},\quad 1<\sigma<2,
\end{equation}

\begin{equation}\label{burg_reg_norm}
\frac{\partial\phi^a}{\partial t}=a\nu_{2}\Delta\phi^a-\kappa\phi^a\frac{\partial\phi^a}{\partial x},\quad\sigma>2,
\end{equation}

\noindent where $\phi(x,t)$ is replaced with $\phi^a(x,t)$ to emphasize its dependence on $a$. These additional viscous terms do not represent the true lowest-order correction, but are rather kept due to the smoothening (regularizing) effect  (see \cite{Droniou03} and references therein) they impose on the solutions of the inviscid Burgers' equation that could otherwise develop discontinuities (shocks) \cite{LighthillWhitham55}. As we show in Sec. IV, they also play an important role in selecting the way in which the system relaxes to the stationary state.

\section{Application to an open system}

We now wish to apply hydrodynamic equations (\ref{burg_reg_frac}) and (\ref{burg_reg_norm}) to a finite system in contact with left and right reservoirs of densities $\rho_{L}=\alpha$ and $\rho_{R}=1-\beta$, respectively. Furthermore, we consider only the stationary limit where $\partial\phi(x,t)/\partial t=0$ so that $\phi(x,t)\rightarrow\phi(x)$. This case was already studied by Monte Carlo simulations with random sequential update \cite{SzavitsUzelac06} in the totally asymmetric case ($p=1$, $q=0$) for various values of $\alpha$ and $\beta$. The obtained density profiles reproduced the phase diagram of the short-range model provided that the current was renormalized by the average hopping length $\lambda_{L}(\sigma)\equiv\zeta_L(\sigma)/\zeta_{L}(\sigma+1)$, where $\zeta_{L}(z)=\sum_{l=1}^{L}l^{-(1+z)}$ is partial sum of the Riemann zeta function $\zeta(z)$ and $L$ is the number of sites on the lattice.

However, a difference was observed at the first-order transition line and in the maximum-current phase. Apart from the localization of a domain-wall at the first-order transition line $\alpha=\beta<1/2$, the exponent that characterizes the algebraic decay of density profile in the maximum-current phase appeared to be $\sigma$-dependent. In particular, the results obtained by Monte Carlo simulations in the maximum-current phase for system of size $L$ showed that a deviation of density from its bulk value $1/2$, $\Delta\rho(n,L)\equiv\left\vert\langle\tau_n\rangle-1/2\right\vert$, obeys the scaling relation

\begin{equation}
\label{mc_scaling}
\Delta\rho(n,L)=L^{-\mu}f(n/L),
\end{equation}

\noindent where $f(x)\sim x^{-\mu}$ for $x\ll 1/2$ and $\mu$ was conjectured to be

\begin{equation}\label{mu}
\mu=min\left\{\frac{\sigma-1}{2},\frac{1}{2}\right\}.
\end{equation}

\noindent In the rest of this Section, we show that this particular dependence on $\sigma$ can be explained within the hydrodynamic approach. Our approach follows the one of Krug \cite{Krug91}, who applied the viscous Burgers' equation (\ref{burg_reg_norm}) in the stationary limit to a finite system of size $l$ in contact with two reservoirs in order to investigate boundary-induced phase transitions in the asymmetric exclusion process with short-range hopping. Imposing the boundary conditions $\rho(0)=\rho_0$ and $\rho(l)=0$, he was then able to show the emergence of characteristic length $\xi$, diverging for $\rho_0$ greater than the critical density $\rho^{*}=1/2$ and finite $\xi\sim(\rho^{*}-\rho_0)^{-1}$ for $\rho_0<\rho^{*}$. Thus the continuous transition occurs between the phase in which $\rho(x)$ decays exponentially to the bulk value $\bar{\rho}\neq\rho^{*}$ and the power-law in which $\xi$ diverges and $\rho(x)$ displays algebraic decay with the exponent $\bar{\mu}=1$, $\rho(x)-\rho^{*}\sim a/x$. This approach gives the correct qualitative picture that explains the second-order phase transition in the asymmetric exclusion process with open boundaries, but the exponent itself is wrong and should be $1/2$ \cite{SchutzDomany93,DerridaEvans93} instead of $1$ \cite{DerridaDomanyMukamel92}. The reason lies in the fact that in general the steady state of an open system is no longer given by the product measure but displays correlations, while these are by default neglected in the mean-field approximation. 

Let us start by assuming that particles jump only in one direction, so that $p=1$ and $q=0$. Due to the finiteness of the system, a distance $1\leq l\leq L$ is now chosen according to the probability $p_l=l^{-(1+\sigma)}/\zeta_L(\sigma+1)$. As it was pointed out in our previous work \cite{SzavitsUzelac06}, the long-range hopping in a finite open system raises the necessity of ``non-local boundary conditions'', in the sense that the exchange of particles with reservoirs now takes place at each site as if there was a bulk reservoir \cite{ParmeggianiFranoschFrey03,ParmeggianiFranoschFrey04,JuhaszSanten04}. As the Fig. 1 shows, one can see that, for example, a particle at site $n$ is removed from the lattice with the probability $\beta_n$, which is the sum of probabilities of all the possible jumps outside the lattice. This includes all the jumps of size $l$ for which $L-n+1\leq l\leq L$, multiplied by the probability $1-\rho_R=\beta$ that the right reservoir is empty

\begin{equation}\label{beta_i}
\beta_{n}=\frac{\beta}{\zeta_{L}(\sigma+1)}\sum_{l=L-n+1}^{L}\frac{1}{l^{\sigma+1}}.
\end{equation}

%
% FIG 1
%

\begin{figure}[!ht]
\centering\includegraphics{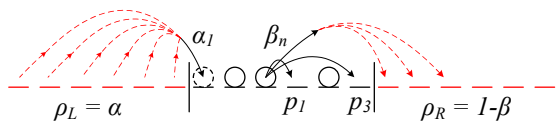}
\caption{(Color online) A schematic picture of an open system with $L=6$ sites. A particle at site $n$ may jump to the right reservoir in a number of ways for which $n+l>L$ (dashed lines), all adding up to the total probability $\beta_n$ to remove the particle from a system. Similarly, a particle may be added from the left reservoir to an empty site $n$ only from those sites in the left reservoir that are within a distance $n\leq l\leq L$ from the site $n$.} 
\label{fig1}
\end{figure}

\noindent Similarly, a particle from the left reservoir is added to an empty site $1\leq n\leq L$ with a probability $\alpha_n$, which is the total probability of all the possible jumps from the left reservoir to an empty site $n$, multiplied by the probability $\rho_L=\alpha$ that the left reservoir is not empty

\begin{equation}\label{alpha_i}
\alpha_{n}=\frac{\alpha}{\zeta_{L}(\sigma+1)}\sum_{l=n}^{L}\frac{1}{l^{\sigma+1}}.
\end{equation}

\noindent A precise meaning of these boundary conditions can be given by noticing that in case of the open boundary conditions, the lattice equations for the average density $\langle\tau_n\rangle(t)$ in the mean-field approximation

\begin{eqnarray}\label{eq_evo_i}
\frac{d}{dt}\langle\tau_n\rangle & = & \alpha_n(1-\langle\tau_n\rangle)+\sum_{m=1}^{n-1}p_{n-m}\langle\tau_m\rangle(1-\langle\tau_n\rangle)-{}\nonumber\\
&& {}-\sum_{m=n+1}^{L}p_{m-n}\langle\tau_n\rangle(1-\langle\tau_m\rangle)-\beta_n\langle\tau_n\rangle,
\end{eqnarray}

\begin{equation}\label{eq_evo_1}
\frac{d}{dt}\langle\tau_1\rangle=\alpha_1(1-\langle\tau_1\rangle)-\sum_{m=2}^{L}p_{m-1}\langle\tau_1\rangle(1-\langle\tau_m\rangle)-\beta_1\langle\tau_1\rangle,
\end{equation}

\begin{equation}\label{eq_evo_L}
\frac{d}{dt}\langle\tau_L\rangle=\alpha_L(1-\langle\tau_L\rangle)+\sum_{m=1}^{L-1}p_{L-m}\langle\tau_m\rangle(1-\langle\tau_L\rangle)-\beta_L\langle\tau_L\rangle,
\end{equation}

\noindent can be written in the same form as in (\ref{tau_n}) and (\ref{K_n}) if we extend the system for additional $L$ sites to the left and the right, but require that $\langle\tau_n\rangle=\alpha$ if $-L<n\leq 0$ and $\langle\tau_n\rangle=1-\beta$ if $L<n\leq 2L$. [A similar reasoning that fixes the values of function in the extended region was used for the numerical solution of the boundary-value problem that involves fractional derivatives \cite{CiesielskiLeszczynski06}.] 

In order to examine the scaling property (\ref{mc_scaling}), we suppose that away from the left boundary (but substantially far away from the right one), density profile $\phi_n=\vert\langle\tau_n\rangle-1/2\vert$ decays algebraically with some unknown exponent $\bar{\mu}>0$, $\phi_n\sim n^{-\bar{\mu}}=a^{\bar{\mu}}/x^{\bar{\mu}}$. A rough estimate of the length scale beyond which this asymptotic behavior sets in is given by $l_{S}\sim a/(\rho_0-1/2)^{1/\bar{\mu}}$, where $\rho_0=\langle\tau_1\rangle$. To employ the hydrodynamic equation for this problem, we assume that the size of the system is large enough that the influence of the right reservoir may be ignored. In that case, we may fix the value of $\phi(x)$ to be equal to $\phi(0)=\rho_0-1/2$ for all $x<0$. Inserting $\phi(x)$ in the Eq. (\ref{burg_reg_frac}) yields the following estimate of various terms for $x>l_{S}$

\begin{equation}\label{burg_left}
-\kappa\phi(x)\frac{\partial\phi(x)}{\partial x}\sim-\bar{\mu}\vert\kappa\vert a^{2\bar{\mu}}x^{-2\bar{\mu}-1},
\end{equation}

\begin{equation}\label{frac_left_sigma}
a^{\sigma-1}\nu_{\sigma}\Delta_{\sigma}\phi(x)\sim \frac{\phi(0)}{\sigma\zeta(\sigma+1)}a^{\sigma-1}x^{-\sigma}+O(a^{\sigma-1+\bar{\mu}}x^{-\sigma-\bar{\mu}}),
\end{equation}

\noindent where the most dominant part of the expression (\ref{frac_left_sigma}) follows from the ``boundary'' condition $\phi(x)=\phi(0)$ for $x<0$. [As the nonlinear term $\phi H_{\sigma}\phi$ is concerned, one can show that it is of the order $O(a^{\sigma-1+\bar{\mu}}x^{-\sigma-\bar{\mu}})$.] Since the left hand side of Eq. (\ref{burg_reg_frac}) must be zero in the stationary regime, terms (\ref{frac_left_sigma}) and (\ref{burg_left}) must be of the same order, which gives the scaling exponent $\bar{\mu}=(\sigma-1)/2$, both in powers of $a$ and $x$.

For $\sigma>2$, the boundary effect due to the fixed value of $\phi(x)$ for $x<0$ requires that one includes both $a\Delta\phi(x)$ and $a^{\sigma-1}\Delta_{\sigma}\phi(x)$ in the equation for $\phi(x)$. This gives the following estimate for $a\Delta\phi(x)$ 

\begin{equation}\label{norm_left_sigma}
a\nu_{2}\Delta\phi(x)\sim\vert\nu_{2}\vert\bar{\mu}(1+\bar{\mu})a^{1+\bar{\mu}}x^{-\bar{\mu}-2},
\end{equation}

\noindent while the expressions (\ref{burg_left}) and (\ref{frac_left_sigma}) remain the same. This leads to equation 2$\bar{\mu}=min\lbrace \sigma-1,1+\bar{\mu}\rbrace$ that has two different solutions depending on the value of $\sigma$: $\bar{\mu}=(\sigma-1)/2$ for $2<\sigma<3$ and $\bar{\mu}=1$ for $\sigma>3$. To summarize, we can write

\begin{equation}\label{mu_mf}
\bar{\mu}=min\left\{\frac{\sigma-1}{2},1\right\}.
\end{equation}

\noindent For $1<\sigma<2$, this result is exactly the same as (\ref{mu}). However, it fails to give the correct value of $\sigma$ for which the short-range regime sets in: according to (\ref{mu_mf}) this value is $\sigma=3$, instead of $\sigma=2$. This failure arises as a result of neglecting the correlations in the mean-field approach, so that for $\sigma>2$ the non-local effect of the boundaries is overestimated.

\subsection{Numerical solution of discrete mean-field equations}

\begin{figure}[!ht]
\centering\includegraphics{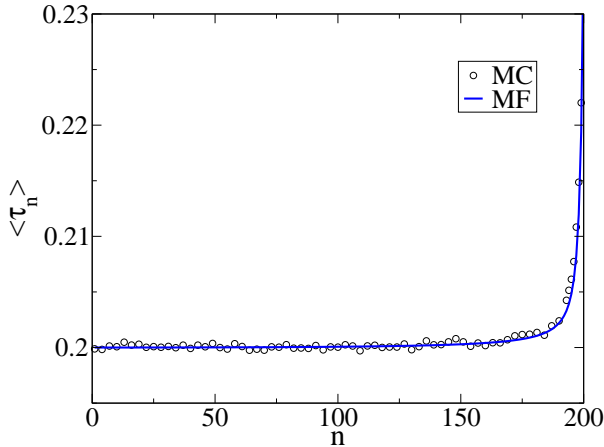}
\caption{(Color online) A comparison of density profiles obtained in the mean-field approximation (line) and by Monte Carlo simulations (symbols) for $\alpha=0.2$, $\beta=0.7$ and $\sigma=1.8$ (low-density phase).}
\label{fig2}
\end{figure}

\begin{figure}[!ht]
\centering\includegraphics{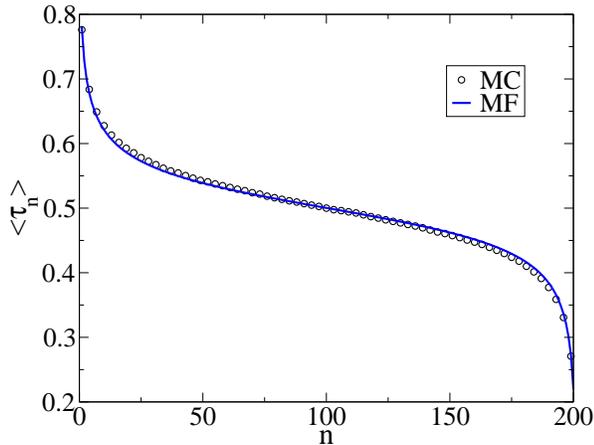}
\caption{(Color online) A comparison of density profiles obtained in the mean-field approximation (line) and by Monte Carlo simulations (symbols) for $\alpha=1.0$, $\beta=1.0$ and $\sigma=1.8$ (maximum-current phase).}
\label{fig3}
\end{figure}

\begin{figure}[!ht]
\centering\includegraphics{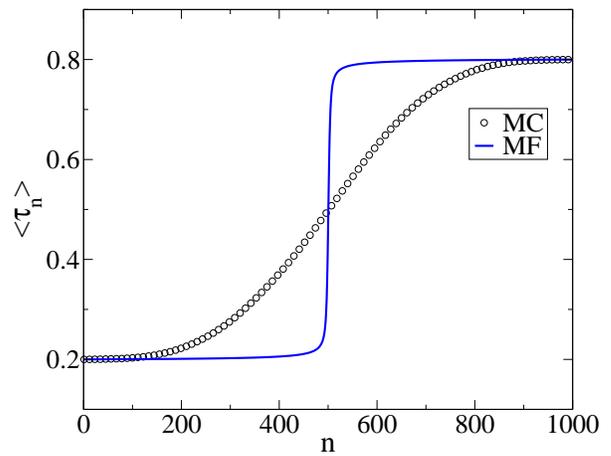}
\caption{(Color online) A comparison of density profiles obtained in the mean-field approximation (line) and by Monte Carlo simulations (symbols) for $\alpha=\beta=0.2$ and $\sigma=1.5$ (coexistence line).}
\label{fig4}
\end{figure}

\begin{figure}[!ht]
\centering\includegraphics{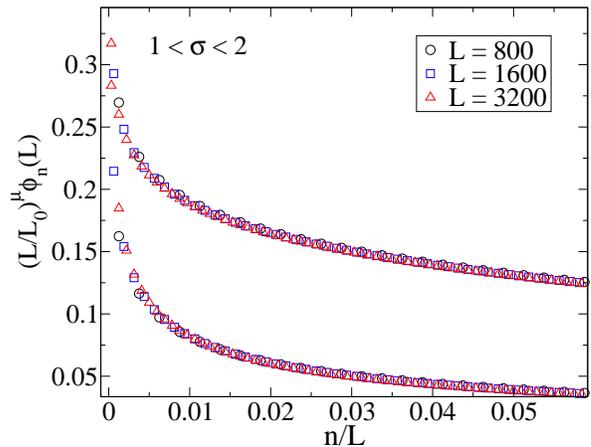}
\caption{(Color online) Deviation $\phi_{n}(L)$ of a density profile from its bulk value $\overline{\rho}=1/2$, obtained from the numerical solution of mean-field equations for various system sizes $L=800$, $1600$, and $3200$ ($\alpha=\beta=1.0$) and for $\sigma=1.2$ and $1.8$ (from top to bottom). The profiles $\phi_{n}(L)$ for the same $\sigma$ are scaled to the profile $\phi_{n}(L_0)$ ($L_0=3200$) according to Eq. (\ref{mc_scaling}) with exponent $\mu=\bar{\mu}$ given by (\ref{mu_mf}).}
\label{fig5}
\end{figure}

\begin{figure}[!ht]
\centering\includegraphics{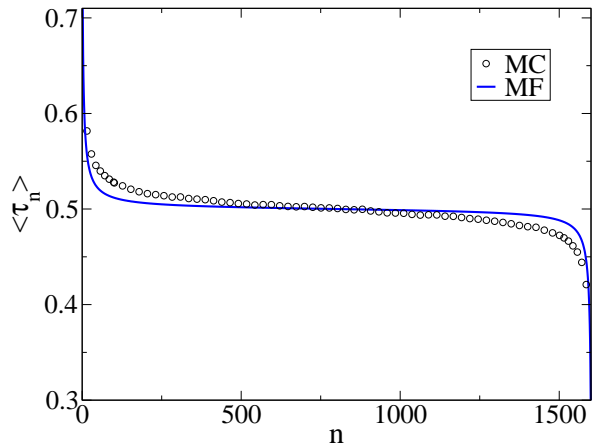}
\caption{(Color online) A comparison of density profiles obtained in the mean-field approximation (line) and by Monte Carlo simulations (symbols) for $\alpha=1.0$, $\beta=1.0$ and $\sigma=2.5$ (maximum-current phase).}
\label{fig6}
\end{figure}

The expression (\ref{mu_mf}) can be checked directly by the numerical solution of the stationary lattice equations (\ref{eq_evo_i})-(\ref{eq_evo_L}) in the mean-field approximation. This case reduces to a problem of finding a zero of a system of $L$ nonlinear equations in $L$ variables, which can be done numerically. For this purpose, we used the HYBRD algorithm taken from the MINPACK library \cite{minpack} for various $\alpha$, $\beta$ and $\sigma$. In the region $1<\sigma<2$, the results reproduce the phase diagram well and the profiles coincide with the results of the Monte Carlo simulations (Fig. \ref{fig2} and \ref{fig3}). The only exception is the line $\alpha=\beta<1/2$ with a sharp domain-wall located in the middle (Fig. \ref{fig4}). This is similar to the mean-field solution in the short-range case \cite{DerridaDomanyMukamel92} which does not take into account fluctuations of the position of the domain wall. These fluctuations for the long-range case \cite{SzavitsUzelac06} can be taken into account using the domain-wall approach in a same manner as it was done for the short-range case with the bulk reservoir \cite{EvansJuhaszSanten03}, where the domain-wall performs random-walk in a potential well. As far as the maximum-current phase is concerned, density profiles satisfy the scaling relation (\ref{mc_scaling}) along with the mean-field exponent $\bar{\mu}$ for all $1<\sigma<2$ (Fig. \ref{fig5}), but the profiles themselves no longer match those obtained in the Monte Carlo simulations as soon as $\sigma>2$ (Fig. \ref{fig6}). This is expected since for $\sigma>2$, $\bar{\mu}$ reads $min\lbrace(\sigma-1)/2,1\rbrace$ while the correct value is $1/2$.

%
% FIG 2
%

%
% FIG 3
%

%
% FIG 4
%

%
% FIG 5
%

%
% FIG 6
%

Let us conclude this Section by giving a simple picture that accounts for the $\sigma$-dependent exponent $\mu$ for $1<\sigma<2$ and its change to the short-range value $1/2$ for $\sigma>2$. First, notice that the nonlinear term $\phi\partial\phi/\partial x$ appears in the hydrodynamic equations no matter what the range of the hopping is. This implies the possibility of reproducing the same $\sigma$-dependent exponent by replacing the long-range hopping with the short-range one, provided that the nearest-neighbors hopping rate is increased by the factor $\lambda(\sigma)$ and that the bulk reservoir is kept with the same $\sigma$-dependent rates $\alpha_n$ and $\beta_n$. Indeed, density profiles obtained by Monte Carlo simulations of such a modified model reproduce the scaling property (\ref{mc_scaling}) with the exponent $(\sigma-1)/2$ for $1<\sigma<2$ and $1/2$ for $\sigma>2$. As far as the large-scale behavior is concerned, the original model is thus reduced to the short-range one with the additional external field that originates from the long-range exchange of particles between the reservoirs and the bulk, while all the other contributions that are result of the long range of hopping are of higher order. [For example, this is the case with the hopping from one site in the bulk to another. As already mentioned earlier in this Section in context of hydrodynamic approach, in the maximum-current phase the nonlocal term $\phi H_{\sigma}\phi$ is of order $O(a^{\sigma-1+\bar{\mu}}x^{-\sigma-\bar{\mu}})$ and does not affect the scaling properties.] Since the external field itself exhibits a power-law dependence in the position on the lattice, it affects the scaling exponent in the maximum-current phase. However, the latter is true only for $1<\sigma<2$, where the influence of the boundaries is stronger than that of the correlations. The boundary value $\sigma=2$  may be reasoned by looking at the average distance $\gamma_L(\sigma)=\sum_n n\cdot\alpha_n/\alpha=\sum_n (L-n+1)\cdot\beta_n/\beta$ at which the particles are created and annihilated \cite{SzavitsUzelac06}. Since $\gamma_L(\sigma)$ diverges with the system size $L$ as $L^{2-\sigma}$ for $1<\sigma<2$ and tends to a finite value for $\sigma>2$, the influence of the external field becomes localized near the boundaries for $\sigma>2$ and the exponent $\mu$ changes to its short-range value $1/2$.

At this point, it is useful to give a more detailed comparison between the present model and the (short-range) model of TASEP with Langmuir kinetics \cite{ParmeggianiFranoschFrey03, ParmeggianiFranoschFrey04}. In the latter case, particles are created and annihilated in the bulk with rates $\Omega_A$ and $\Omega_D$, respectively, that depend on the system size as $L^{-1}$. In the more general case \cite{JuhaszSanten04}, these rates have been extended to $\Omega_A,\Omega_D\sim L^{-a}$, $1<a<2$, which is the same power-law dependence displayed by $\alpha_n$ and $\beta_n$ \cite{SzavitsUzelac06} with $a=\sigma$. As a result of this dependence, both models exhibit localization of the domain-wall at the coexistence line $\alpha=\beta<1/2$ for $1<a<2$, where the width of the domain-wall $L^{-a/2}$ is determined \emph{only} by the $a$-dependent potential well in which the domain-wall performs a random walk \cite{EvansJuhaszSanten03, JuhaszSanten04,SzavitsUzelac06}. On the other hand, the distinction between these two models becomes pronounced in the maximum-current phase, where the fact that the rates $\alpha_n$ and $\beta_n$ depend on position on the lattice becomes important. This is in contrast to the short-range case with bulk reservoir where the creation and annihilation of particles with homogeneous rates does not affect the scaling exponent which remains $1/2$ for all $1<\sigma<2$, but merely determines the characteristic time that particles spend on the lattice \cite{JuhaszSanten04}.

\section{Dynamical scaling}

Another issue we wish to address here concerns the dynamical-scaling properties of the late-time approach to the stationary state. In the short-range exclusion process, the longest relaxation time $\tau$ scales with a system size $L$ as $\tau\sim L^z$ where $z$ equals $2$ for the symmetric ($p=q$) \cite{deGierEssler06} and $3/2$ for the asymmetric case ($p\neq q$) \cite{GwaSpohn92,Kim95,deGierEssler05,deGierEssler06}. These exponents were brought into connection with dynamical exponents of the corresponding hydrodynamic equations with additional noise term. In particular, dynamical exponents $z=2$ and $z=3/2$ correspond to those of the noisy Edwards-Wilkinson (EW) \cite{EdwardsWilkinson82} and the noisy Burgers' equation \cite{ForsterNelsonStephen77}, respectively, where the latter can be mapped to the Kardar-Parisi-Zhang (KPZ) equation \cite{KardarParisiZhang86} of surface growth.

Generally, the late-time characteristics of a system can be probed by looking at the two-point autocorrelation function which in a translationally invariant system takes the form

\begin{equation}
C(x,t)\equiv\langle\phi(0,0)\phi(x,t)\rangle,
\end{equation}

\noindent where the averaging $\langle\dots\rangle$ is taken over the noise histories. For processes described by EW or KPZ equations, $C(x,t)$ is known to give the following scaling relation,

\begin{equation}\label{cf}
C(x,t)=x^{2\chi-2}F(t/x^z),
\end{equation}

\noindent where $\chi$ denotes the roughening exponent and is equal to $1/2$ in both cases \cite{Krug97,ForsterNelsonStephen77}. 

\indent Both EW and KPZ equations have been generalized by replacing the usual diffusion term with the fractional one \cite{MannWoyczynski01, Katzav03}. This gives the fractional EW equation

\begin{equation}\label{noisy_ew_frac}
\frac{\partial h}{\partial t}=\nu\Delta_{\sigma}h+\eta,\quad 0<\sigma\leq 2,
\end{equation}

\noindent where $\phi(x,t)=\partial h(x,t)/\partial x$ and $\eta(x,t)$ is the noise. A simple dimensional analysis gives the values of exponents $\chi=(\sigma-1)/2$ and $z=\sigma$ \cite{MannWoyczynski01}. On the other hand, the fractional KPZ equation,

\begin{equation}\label{noisy_kpz_frac}
\frac{\partial h}{\partial t}=\nu\Delta_{\sigma}h+\frac{\lambda}{2}\left(\frac{\partial h}{\partial x}\right)^2+
\eta,\quad 0<\sigma\leq 2,
\end{equation}

\noindent displays a more complex behavior with various values of $\chi$ and $z$ depending on the value of $\sigma$ and whether the spatial correlations of the noise are relevant or not \cite{Katzav03}. Particularly, in case these correlations are irrelevant, there is a weak-coupling regime for $\sigma<3/2$ with the same critical exponents as in the fractional EW equation (\ref{noisy_ew_frac}), i.e. $\chi=(\sigma-1)/2$ and $z=\sigma$, and a strong-coupling regime for $\sigma>3/2$ with the usual KPZ exponents $\chi=1/2$ and $z=3/2$.

%
% FIG 7
%

\begin{figure}[!ht]
\centering\includegraphics{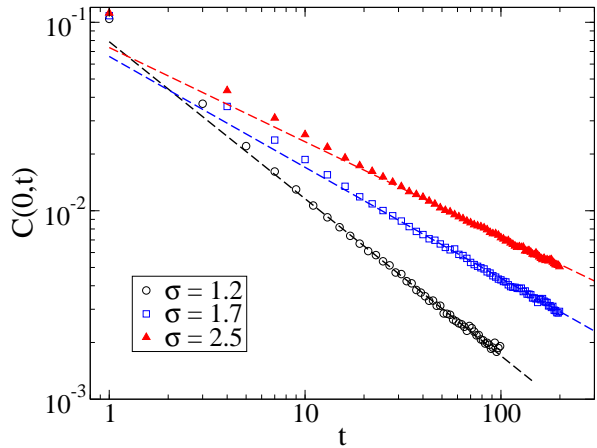}
\caption{(Color online) Time-decay of the two-point autocorrelation function $C(0,t)\sim t^{-1/z}$ for the symmetric exclusion process with long-range hopping on the half-filled periodic lattice ($L=10^4$, averaged over $10^7$ independent MC runs). Dashed lines refer to the expected value of $z=min\lbrace\sigma,2\rbrace$.}
\label{fig7}
\end{figure}

%
% FIG 8
%

\begin{figure}[!ht]
\centering\includegraphics{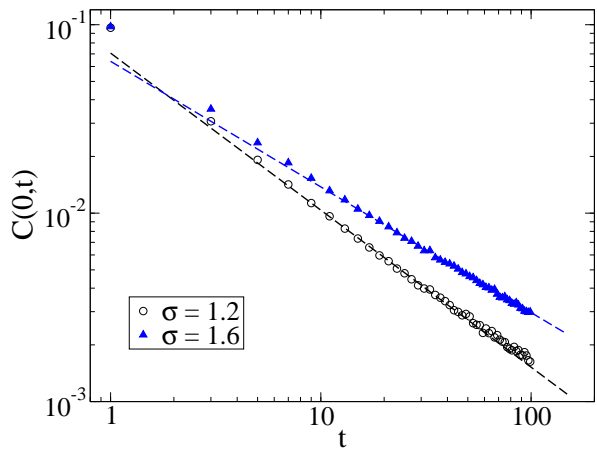}
\caption{(Color online) Time-decay of the two-point autocorrelation function $C(0,t)\sim t^{-1/z}$ for the asymmetric exclusion process with long-range hopping on the half-filled periodic lattice ($L=10^4$, averaged over $10^7$ independent MC runs). Dashed lines refer to the expected value of $z=min\lbrace\sigma,3/2\rbrace$.}
\label{fig8}
\end{figure}

\indent To compare this picture with the late-time approach to the stationary state of the exclusion process with long-range hopping, we computed the corresponding two-point autocorrelation function $C(i-j,t)$ using Monte Carlo simulations in a system with periodic boundary conditions. The averaging was taken over $10^7$ independent runs with $L=10^4$ for times up to $t=100$ Monte Carlo steps/site. Dynamical exponent $z$ was then extracted from the time-decay of the peak $C(0,t)\sim t^{-1/z}$ \cite{PierobonParmeggianiOppenFrey05}. The results for various $\sigma$ are presented in Fig. \ref{fig7} and Fig. \ref{fig8} for the symmetric and the asymmetric case, respectively. In the symmetric case, dynamical exponent is in a good agreement with $z=min\lbrace\sigma,2\rbrace$ predicted by the fractional ($1<\sigma<2$) and the normal ($\sigma>2$) EW equation. In the asymmetric case, results presented in Fig. \ref{fig8} confirm a change in the exponent at $\sigma=3/2$ from the weak-coupling to the strong-coupling regime according to the predicted value $z=min\lbrace\sigma,3/2\rbrace$ of the fractional KPZ equation with irrelevant spatial correlations of the noise. In the interpretation of the Ref. \cite{Katzav03}, this change reflects a tendency of a system to relax through its fastest ``component'' that has a lower value of $z$.

\section{Conclusion}

In this work we studied the large-scale dynamical and stationary properties of the exclusion process with long-range hopping where each particle may jump any distance $l\geq 1$ with the probability that decays with $l$ as $p_l\sim l^{-(1+\sigma)}$. We started on an infinite lattice with no boundary conditions present and calculated the continuous limit of lattice equations in the Fourier space \cite{Tarasov06}. For the symmetric hopping this yields the same result as in Ref. \cite{Jara07} in which the fractional diffusion equation replaces the usual one for $1<\sigma<2$. For the asymmetric hopping, we adopted the non-rigorous approach that consists of decoupling lattice equations by means of the mean-field approximation. As in the short-range case, the result is still given by the inviscid Burgers' equation, but the particle current $j(\rho)=(p-q)\lambda(\sigma)\rho(1-\rho)$ has increased by the factor $\lambda(\sigma)=\zeta(\sigma)/\zeta(\sigma+1)$. A true signature of the long range of hopping was found by inspecting the lowest-order corrections in powers of $a$ which in this case correspond to the fractional Laplacian for $1<\sigma<2$ and the usual one for $\sigma>2$. If these terms are kept, the equation becomes equivalent to the deterministic part of the fractional ($1<\sigma<2$) \cite{Katzav03} and the original KPZ equation ($\sigma>2$) \cite{KardarParisiZhang86}. To check this connection we analyzed the late-time approach to the stationary state in a finite system with periodic boundary conditions. For that purpose we extracted dynamical exponent from the time-decay of the autocorrelation function obtained from Monte Carlo simulations and compared it to the dynamical exponent $z=min\lbrace\sigma,3/2\rbrace$ of the fractional KPZ equation \cite{Katzav03}. A very good agreement of these two exponents lead us to the conclusion that the system always relaxes through its fastest ``component'' that has the lower value of $z$ \cite{Katzav03}. Similar analysis was also carried out for the symmetric case yielding a very good agreement with the known exponents $z=\sigma$ and $z=2$ of the fractional ($1<\sigma<2$) and the usual Edwards-Wilkinson equation ($\sigma>2$), respectively.

In case of the open boundary conditions, the above approach was applied only phenomenologically. In addition, the long range of hopping in a finite system introduces ``non-local boundary conditions'' by means of the inhomogeneous external field that creates and annihilates particles at each site. In the maximum-current phase, this causes the change in the exponent $\mu$ that determines the algebraic decay of density profile. The resulting exponent obtained in the mean-field approximation justifies the earlier conjecture \cite{SzavitsUzelac06} that $\mu=(\sigma-1)/2$ for $1<\sigma<2$. We argue that the short-range regime sets in immediately after, for $\sigma=2$.

\appendix*
\section{Riesz fractional derivative}

Consider the left- and the right-hand Riemann-Liouville fractional derivatives $_a\mathcal{D}_{x}^{\sigma}$ and $_x\mathcal{D}_{b}^{\sigma}$ of order $\sigma$, respectively, whose action on the suitable function $f(x)$ is defined as \cite{SamkoKilbasMarichev93}

\begin{equation}\label{rl_plus}
_a\mathcal{D}^{\sigma}_{x}f(x)=\frac{1}{\Gamma(n-\sigma)}\frac{d^n}{dx^n}\int_{a}^{x}f(\xi)(x-\xi)^{n-\sigma-1},
\end{equation} 

\begin{equation}\label{rl_minus}
_x\mathcal{D}^{\sigma}_{b}f(x)=\frac{(-1)^n}{\Gamma(n-\sigma)}\frac{d^n}{dx^n}\int_{x}^{b}f(\xi)(\xi-x)^{n-\sigma-1},
\end{equation}

\noindent where $n$ is the smallest integer exceeding $\sigma$. For $a=-\infty$ and $b=\infty$, integrals (\ref{rl_plus}) and (\ref{rl_minus}) have the following simple property with respect to the Fourier transform $\mathcal{F}$,

\begin{equation}\label{weyl_left_fourier}
\mathcal{F}\lbrace _{-\infty}\mathcal{D}_{x}^{\sigma}f(x)\rbrace=(-ik)^{\sigma}\hat{f}(k),
\end{equation}

\begin{equation}\label{weyl_right_fourier}
\mathcal{F}\lbrace _x\mathcal{D}_{\infty}^{\sigma}f(x)\rbrace=(ik)^{\sigma}\hat{f}(k),
\end{equation}  

\noindent where $\hat{f}(k)=\mathcal{F}\lbrace f(x)\rbrace$. If we now take the linear combination of $_{-\infty}\mathcal{D}_{x}^{\sigma}$ and $_x\mathcal{D}_{\infty}^{\sigma}$ given by

\begin{equation}\label{Delta}
\Delta_{\sigma}f(x)\equiv-\frac{_{-\infty}\mathcal{D}_{x}^{\sigma}+{}_x\mathcal{D}_{\infty}^{\sigma}}{2cos(\pi\sigma/2)},
\end{equation}

\noindent we get the Riesz fractional derivative (also known as fractional Laplacian) with the following property 

\begin{equation}
\mathcal{F}\lbrace\Delta_{\sigma}f(x)\rbrace=-\vert k\vert^{\sigma}\hat{f}(k).
\end{equation}

\noindent On the other hand, the first term in (\ref{B2}) corresponds to the following linear combination of $_{-\infty}\mathcal{D}_{x}^{\sigma}$ and $_x\mathcal{D}_{\infty}^{\sigma}$

\begin{equation}\label{H}
H_{\sigma}f(x)\equiv\frac{_{-\infty}\mathcal{D}_{x}^{\sigma}-{}_x\mathcal{D}_{\infty}^{\sigma}}{2sin(\pi\sigma/2)},
\end{equation}

\noindent with the following property

\begin{equation}
\mathcal{F}\lbrace H_{\sigma}f(x)\rbrace=-i sgn(k)\vert k\vert^{\sigma}\hat{f}(k). 
\end{equation}

\noindent Both $\Delta_{\sigma}$ and $H_{\sigma}$ are special cases of more general fractional derivative defined as the inverse of the Feller potential (see \cite{SamkoKilbasMarichev93} for details).\\

\begin{acknowledgments}
This work was supported by the Croatian Ministry of Science, Education and Sports through grant No. 035-0000000-3187.
\end{acknowledgments}

\end{document}